\documentclass[aps,pra,twocolumn]{revtex4}
\usepackage{graphicx}
\usepackage{dcolumn}
\usepackage{bm}

\newcommand{\ke}[1]{|#1\rangle}

\newcommand{\da}{^\dagger}
\newcommand{\pt}[1]{\left( #1 \right)}

\begin{document}

\title{Collective effects in the dynamics of driven atoms in a high-Q resonator}

\author{Stefano Zippilli$^1$}
\author{Giovanna Morigi$^1$}\author{Helmut Ritsch$^2$}
\affiliation{$^1$ Abteilung f\"ur Quantenphysik,
Universit\"at Ulm, D-89069 Ulm, Germany\\
$^2$ Institut f\"ur Theoretische Physik,
Universit\"at Innsbruck, A-6020 Innsbruck, Austria}
\date{\today}
\begin{abstract}
We study the quantum dynamics of N coherently driven two-level atoms coupled
to an optical resonator. In the strong coupling regime the cavity field
generated by atomic scattering interferes destructively with the pump on the
atoms. This suppresses atomic excitation and even for strong driving fields
prevents atomic saturation, while the stationary intracavity field amplitude
is almost independent of the atom number. The magnitude of the interference
effect depends on the detuning between laser and cavity field and on
the relative atomic positions and is strongest for a
wavelength spaced lattice of atoms placed at the antinodes of the cavity
mode. In this case three dimensional intensity minima are created in the
vicinity of each atom. In this regime spontaneous emission is suppressed and
the dominant loss channel is cavity decay. Even for a cavity linewidth larger
than the atomic natural width, one regains strong interference through the
cooperative action of a sufficiently large number of atoms. These results give
a new key to understand recent experiments on collective cavity cooling and
may allow to implement fast tailored atom-atom interactions as well as
nonperturbative particle detection with very small energy transfer.
\end{abstract}

\maketitle

\section{Introduction}

Cavity quantum electrodynamics CQED using optical resonators has experienced
important experimental progress in recent years. Several experimental groups
achieved remarkable milestones in the realization of well defined strongly
coupled atom-field systems~\cite{Pinkse04,Hood00,Pinkse00,Mundt02,Sauer03,KimbleFORT03,Kimble95,Kuhn02,Kimble03a,MPQ01,Horak02,Zimmerman03,Hemmerich03,Chan03,Black03,Kimble03b}.
This has lead to numerous applications like single atom trapping by a single
photon~\cite{Pinkse04,Hood00,Pinkse00}, conditional quantum phase shifts of very weak
fields~\cite{Kimble95}, deterministic sources of entangled
photons~\cite{Kuhn02,Kimble03a} and a single atom thresholdless
laser~\cite{Kimble03b}.  Direct observations of the field mode structure~\cite{MPQ01,Horak02,Eschner01} and of the
mechanical effects of the cavity field on the atomic motion~\cite{Hood00,Pinkse00,Juergen03} have been reported
culminating in the recent demonstration of cavity induced cooling of single trapped atoms~\cite{Pinkse04}.

Renewed interest was devoted to large ensembles of atoms commonly coupled to a cavity
field with several modes, where collective atomic effects play a central role in the coupled atom-field
dynamics~\cite{Zimmerman03,Hemmerich03,Chan03,Black03}.
These experiments have lead to unexpected results
and opened new theoretical questions. As a particular example, collective
phenomena in the presence of an external transverse driving field involving
many atoms at different positions are not fully
understood~\cite{Chan03,Black03}.

In this work we study theoretically the dynamics of coherently driven atoms in
a resonator. Our investigation takes into account the atomic spontaneous
emission, the finite transmittivity of the cavity mirrors and the spatial
structure of the cavity mode. The system state is characterized by the atomic
fluorescence rate and the signal at the cavity output and studied as a function
of the system's parameter. To get further information on the system, we
calculate the probe absorption spectrum and the field distribution in the
vicinity of the atoms.

Our results extend the studies of~\cite{Zippilli04a}. There we showed that in the strong coupling regime the system dynamics
exhibit enhanced cavity emission accompanied by suppression of fluorescence
which occurs, for more than one atom, when the atoms are spatially localized
such that they emit in phase into the cavity mode. This phenomenon shares
several analogies with the behaviour found in the case of a single atom inside
a lossless resonator~\cite{Alsing92,Alsing92a}. In fact, this behaviour can be
traced back to destructive interference between the laser and the cavity field
generated by atomic scattering, such that the atoms couple to a vanishing
electric field. As a consequence, we show that the stationary cavity field is
independent of the number of atoms and cavity decay becomes the dominant
channel of dissipation. In a good cavity this allows to measure the light
dissipated through cavity decay without destroying the interference, which is
vital if one wants to get information on the cavity field and hence the current
atomic positions. In addition, when the strong coupling regime is achieved by a
large number of ordered atoms the dynamics we find are consistent with the
experimental observations by Vuletic and coworkers~\cite{Chan03,Black03}.

This article is organized as follows. In section II the model is introduced.
In section III we review the results obtained for a lossless
resonator~\cite{Alsing92} and investigate the system's dynamics when the decay
rate of the cavity is finite. In section IV the field inside the cavity is
investigated by means of (i) an additional weak laser coupled to the atom and
(ii) an additional atom weakly coupling to the cavity field. In section V the
scaling of the system dynamics with the number of atoms is studied and the
results are discussed in connection with the experimental observations
in~\cite{Chan03,Black03}. In section VI the results are summarized and
discussed, and several outlooks are provided.
The appendices report the details of the calculations presented in
Secs.\ III and IV.

\section{The Model}

\begin{figure}[h]
\includegraphics[width=4.5cm]{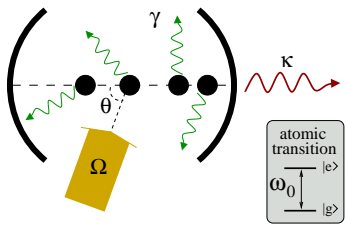}
\caption{N atoms couple to a 1D optical resonator and are driven
transversally by a laser which illuminate them homogeneously. The inset shows
the atomic transition which is relevant to the dynamics.}
\label{Fig:1}
\end{figure}

We consider $N$ identical and point-like atoms, whose dipole transitions
couple resonantly with the standing-wave mode of an optical resonator.
The atoms are assumed to be located along the axis of
the resonator, which we denote with the $x$-axis, and their
center--of--mass motion is neglected. The relevant atomic degrees of freedom
are the ground and excited electronic states $|g\rangle$, $|e\rangle$ of the
dipole transition, which is at frequency $\omega_0$. The transition couples to
the resonator's mode at frequency $\omega_c$ and wave vector $k$,
and it is driven by a laser at
frequency $\omega_{\rm L}$ and Rabi frequency $\Omega$, as shown in
Fig.~\ref{Fig:1}. The laser is assumed to be a classical field.
The dynamics of the composite system is described by the
master equation for the density matrix $\rho$ of atoms and cavity mode
\begin{eqnarray}
\label{Master:0}
  \frac{\partial}{\partial t}\rho =\frac{1}{{\rm
      i}\hbar}[H,\rho]+{\cal L}\rho
  + {\cal K}\rho
\end{eqnarray}
where $H$ is the Hamiltonian for the coherent dynamics,
and ${\cal L}$, ${\cal K}$ are the
superoperators describing dissipation due to spontaneous decay and cavity losses. In the reference frame
rotating at the laser frequency $\omega_{\rm L}$ the Hamiltonian $H$ has
the form
\begin{eqnarray}
\label{H}
H
&=&-\hbar\delta_ca^\dagger a-\hbar
\Delta\sum_{n=1}^N |e\rangle_n\langle e|\\
&+&\hbar\sum_{n=1}^N \left[g(x_n)(a\sigma_n^{\dagger} +a^\dagger
    \sigma_n)+\Omega\left({\rm
      e}^{i\phi_n}\sigma_n^{\dagger}+{\rm
      e}^{-i\phi_n}\sigma_n\right)\right]\nonumber
\end{eqnarray}
where $a$, $a^\dagger$ are the annihilation and creation operators of a cavity
photon; $\sigma_n=|g\rangle_n\langle e|$,
$\sigma_n^{\dagger}=|e\rangle_n\langle g|$, are the dipole operators for the
atom at the position $x_n$;
$\delta_c=\omega_{\rm L}-\omega_c$, $\Delta=\omega_{\rm L}-\omega_0$
are the detunings of the laser from the frequency of the cavity and of the
dipole, respectively. The coupling constant between the dipole at position
$x_n$ and the cavity mode is
$g(x_n)=g_0\cos kx_n$, while the coupling with the driving laser
depends on the atomic position through the phase $\phi_n=kx_n\cos \theta$,
where $\theta$ is the angle between the cavity axis and the
propagation direction of the laser.
Finally, the incoherent dynamics is described by the superoperators
\begin{eqnarray}
{\cal L}\rho&=&\frac{\gamma}{2}\sum_n\left( 2 \sigma_n\rho \sigma_n^{\dagger}-
    \sigma_n^{\dagger} \sigma_n\rho-\rho \sigma_n^{\dagger} \sigma_n\right)\\
{\cal
    K}\rho&=&\frac{\kappa}{2}(2a\rho a^\dagger -a^\dagger a\rho-\rho a^\dagger
  a) \end{eqnarray}
where $\gamma$ is the rate of spontaneous emission of the dipole into the modes
that are external to the cavity and $\kappa$ is the cavity decay
    rate. Collective effects in the spontaneous decay are neglected here, as
    the average distance between the atoms is assumed to be of the order of
    several wavelengths.

\section{Enhanced cavity emission in high-finesse cavities}

In~\cite{Alsing92} it has been shown that the steady state of a lossless
cavity, coupled to a dipole and driven transversally, is a pure state, such
that the energy of the atom and cavity mode is conserved. This regime is
accessed when the driving laser is resonant with the cavity mode. Then, the
atom is in the ground state and the cavity mode field, generated by atomic
scattering, is described by a coherent state whose amplitude is determined by
the intensity of the laser. This can be easily seen in Eq.~(\ref{Master:0}) for
$\kappa=0$ and $N=1$, after moving to the reference frame described by the
unitary transformation
\begin{equation}
\label{Displace} {\cal D}(\beta)=\exp(\beta a^{\dagger}-\beta^*a),
\end{equation}
which corresponds to displacing the field inside the cavity by the
amplitude $\beta=-\Omega{\rm  e}^{{\rm i}\phi}/\bar{g}$ (here,
$\bar{g}=g(x)\neq 0$~\cite{Footnote:g=0}).
In this reference frame and for $\delta_c=0$, Eq.~(\ref{H}) takes the form
of the Jaynes-Cummings Hamiltonian and the steady state of the transformed
master equation (\ref{Master:0}) is evidently
the state $|g,0\rangle$. In the original reference frame this corresponds to
the steady state
\begin{equation}
\label{Steady}
\rho_{\rm ss}=|g,\beta\rangle
\langle g,\beta|
\end{equation}
which is a pure state. In fact, the state $|g,\beta\rangle$ is
eigenstate of $H$ and is stable for $\kappa=0$. In particular,
although the atom is driven both by laser and cavity mode, it is in the ground
state as a result of the destructive interference between the atomic
excitations induced by the two fields. Therefore, there is no atomic
fluorescence and consequentely
the only dissipation channel of the system is closed.
State~(\ref{Steady}) exhibits thus the characteristic of a dark
state. Moreover, as the atom does not scatter cavity photons and the cavity is
assumed to be lossless, the energy of the cavity field is conserved.

In any realistic setup an optical resonator has a finite decay rate $\kappa$.
When the atom is driven and $\delta_c=0$, the field decay induces dephasing at
the atomic position and the state (\ref{Steady}) has a finite
lifetime. However, it is reasonable to expect
Eq.~(\ref{Steady}) to approximate the steady state for $\kappa$
sufficiently small. In the following, we investigate the
intensity of the fluorescence signal and of the signal
at the cavity output in different parameter regimes, thereby
verifying under which conditions energy dissipation through spontaneous decay
can be neglected and when the state inside the cavity can be approximated by
a coherent state.

In this section we restrict to the case of one atom
and take $g(x)=\bar{g}\neq 0$. The density matrix at time $t$
for $\kappa\neq 0$ can be analytically determined by means of a
perturbative expansion in the small parameter $\kappa$, where cavity decay is
assumed to be slower than the rate at which the atom reaches the steady
state~\cite{Footnote:1}. At this purpose, we rewrite
the master equation (\ref{Master:0}) in the form
\begin{equation}
\label{Master:Unravel} \frac{\partial}{\partial t}\rho= \frac{1}{{\rm
i}\hbar}\left(H_{\rm eff}\rho-\rho H_{\rm eff}^{\dagger}\right) + J\rho+\kappa
{\cal K}_0\rho
\end{equation}
where
\begin{eqnarray}
\label{H:eff}
H_{\rm eff}
&=&\hbar \bar{g}(a\sigma^{\dagger} +a\da \sigma)+
\hbar\Omega(\sigma^{\dagger} +\sigma)\\
& &-\hbar\left(\Delta+{\rm i}\frac{\gamma}{2}\right)|e\rangle\langle e|
-{\rm i}\hbar \frac{\kappa}{2} a\da a\nonumber
\end{eqnarray}
and $J\rho=\gamma\sigma \rho\sigma^{\dagger}$, ${\cal K}_0\rho=a\rho
a^{\dagger}$ are the jump operators. The formal solution of
Eq.~(\ref{Master:Unravel}) is~\cite{Carmichael}
\begin{equation}
\label{Sol:1} \rho(t)={\cal S}(t)\rho(0)+\int_0^t{\rm d}\tau {\cal
S}(t-\tau)(J+\kappa {\cal K}_0)\rho(\tau)
\end{equation}
with ${\cal S}(t)\rho(0)=\exp(-{\rm i}H_{\rm eff}t/\hbar)\rho(0) \exp({\rm
i}H_{\rm eff}^{\dagger}t/\hbar)$. The perturbative expansion of
Eq.~(\ref{Sol:1}) at second order in $\kappa $ is reported in Appendix A.
The photon scattering rate at time $t$ by the atom into the
modes of the continuum is $I_{\rm at}=\gamma {\rm
  Tr}\{\sigma^{\dagger}\sigma \rho(t)\}$, and grows quadratically with $\kappa$. At the cavity
ouput the rate of photon scattering $I_{\rm cav}=\kappa {\rm Tr}\{a\da a
\rho(t)\}$ at time $t$ and in lowest order is linear in $\kappa$
and $I_{\rm cav}\gg I_{\rm at}$. An
instructive case is found in the limit $\Delta=0$ and $\bar{g}\gg\gamma$.
Here, these
expressions acquire the simple form
\begin{eqnarray}
\label{Res:1}
&&I_{\rm at}\approx
\kappa\frac{\Omega^2}{\bar{g}^2}\frac{1}{2C_1}\\
&&I_{\rm cav}\approx
\kappa\frac{\Omega^2}{\bar{g}^2}\left(1-\frac{1}{2C_1}\right)
\label{Res:2}
\end{eqnarray}
where $C_1=2\bar{g}^2/\gamma\kappa$ is the cooperativity parameter per
atom~\cite{Kimble94}. Thus, the two signals depend on $\kappa$ through the
cooperativity parameter $C_1$ and the factor $\kappa\Omega^2/\bar{g}^2$, which
is the decay rate of a cavity with mean photon number $\langle n\rangle
=\Omega^2/\bar{g}^2$. From Eqs.~(\ref{Res:1}) and~(\ref{Res:2}) it is visible that
 for $C_1\gg 1$ the fluorescence signal is orders of
magnitude smaller than the intensity at the cavity output.

\begin{figure}[h]
\includegraphics[width=7cm]{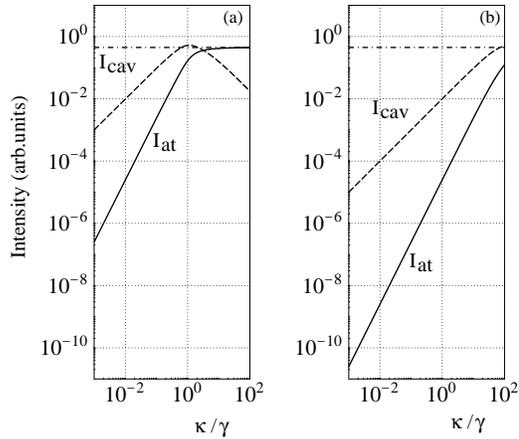}
\caption{$I_{\rm cav}$ (dashed line) and $I_{\rm at}$ (solid line)
as a function of $\kappa$ in units of $\gamma$. Here, $\Omega=\gamma$,
$\Delta=\delta_c=0$ and (a) $g(x)=\gamma$, (b) $g(x)=10\gamma$.
The orizontal dashed-dotted line gives the rate of fluorescence of the atom in
free space.}
\label{Fig:Compare:2}
\end{figure}

Figure~\ref{Fig:Compare:2} displays $I_{\rm at}$ and $I_{\rm cav}$ as a
function of $\kappa$ and for two different values of $\bar{g}$. The curves have
been calculated by solving numerically~(\ref{Master:Unravel}). Here, for a
wide range of values of the cavity decay rate $I_{\rm cav}$ exhibits a
linear behaviour as a function of $\kappa$, while
$I_{\rm at}$ is quadratic. Moreover, when $\bar{g}$
is increased (and thus when the cooperativity parameter is increased)
the relation $I_{\rm at}\ll I_{\rm cav}$ is fulfilled for a wider range of
values of $\kappa$, as it
is visible by comparing Fig~\ref{Fig:Compare:2}(a) with
Fig~\ref{Fig:Compare:2}(b). In particular, for $\bar{g}=10\gamma$
the signal $I_{\rm cav}$ largely exceeds the fluorescence
signal $I_{\rm at}$ even for $\kappa>\gamma$.

\begin{figure}[h]
\includegraphics[width=9.5cm]{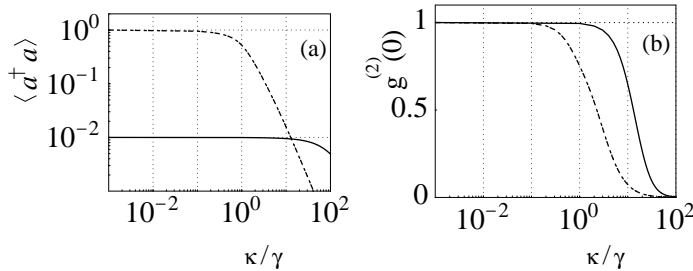}
\caption{(a) Mean number of cavity photon and (b)
second-order correlation function $g^{(2)}(0)$ as a function of
$\kappa$ in units of $\gamma$. Here, $\Delta=\delta_c=0$, $\Omega=\gamma$ and
$g(x)=\gamma$ (dashed line), $g(x)=10\gamma$ (solid line).}
\label{Fig:g2}
\end{figure}

The zero-time correlation function $g^{(2)}(0)$ of the signal
at the cavity mirror gives further insight into the dynamics of the cavity
field. Figure~\ref{Fig:g2}(b) displays $g^{(2)}(0)$ as a
function of $\kappa$ and Fig.~\ref{Fig:g2}(a) displays the corresponding
average number of cavity photons $\langle a^{\dagger}a\rangle$. From
these figures one sees that
the cavity field exhibits a Poissonian behaviour for a fairly wide range of
values of $\kappa$, corresponding to large cooperativity parameters. This
behaviour is verified even when the average number of
cavity photons is very small (solid line in Fig.~\ref{Fig:g2}(a) and (b)). It
shows that the cavity mode is in a coherent state,
independently of the average energy of the cavity field. This behaviour
contrasts dramatically with the antibunching observed
when the pump is set directly on the cavity~\cite{Kimble94,Brecha99}.

\section{Probing the system}

In this section we investigate the response of the system to a weak probe
in the parameter regime for which the steady state
is given to good approximation by state~(\ref{Steady}).
We restrict to the case of one atom,
whose dipole transition couples to the driving field and to the cavity mode,
and consider the spatial dependence of the coupling.
The resonator's
mode function is a standing wave with $g(x)=g_0\cos kx$ and the
atom is assumed to be at position $x$ such that $g(x)=\bar{g}\neq 0$.
The laser is a plane
wave, and its phase $\phi_n$ depends on the atomic position $x_n$ through the
relation $\phi(x)=kx\cos\theta$, where $\theta$ is the angle between the
direction of propagation of the laser and the cavity axis. We assume that laser
and cavity are resonant, and analyze the system's response to two types of
probe: (i) a weak laser field, coupling to the atomic dipole, as shown in
Fig.~\ref{Fig:2}; (ii) a
second atom of a different species, whose dipole transition frequency is
far--off resonance from the cavity frequency, thereby negligibly perturbing
the system.

\subsection{Excitation spectrum}

\begin{figure}[h]
\includegraphics[width=8cm]{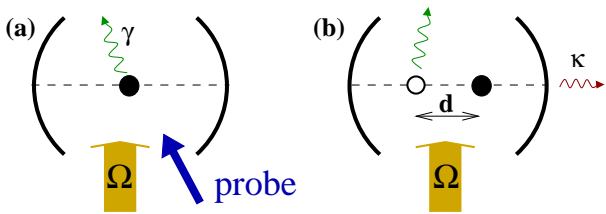}
\caption{(a) Excitation spectrum: A weak probe is coupled to the atomic dipole and
  its frequency is scanned through atomic resonance. The fluorescence signal
  is measured as a function of the probe detuning. (b) A second atom of another species,
weakly coupled to cavity and pump fields, probes the cavity field inducing a
position-dependent phase shift on the cavity field. The phase shift can be
measured by means of homodyne detection or by measuring the fluorescence of the
second atom.} \label{Fig:2}
\end{figure}

\begin{figure}[h]
\includegraphics[width=8cm]{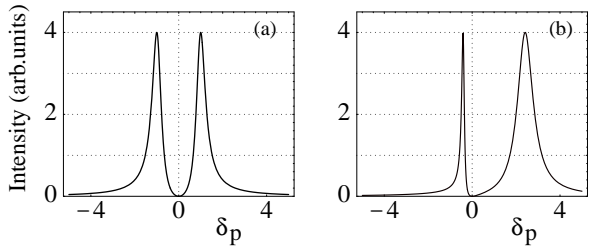}
\caption{Excitation spectrum $w(\delta_P)$
as a function of the probe detuning $\delta_P$ in units of $\gamma$,
for the parameters $g(x)=\gamma$, $\delta_c=0$,
$\kappa=0$ and (a) $\Delta=0$, (b) $\Delta=-2\gamma$.}
\label{Fig:Probe}
\end{figure}

We consider a probe driving the atom
as illustrated in Fig.~\ref{Fig:2}(a) and evaluate the excitation spectrum,
namely the rate of photon scattering into the modes external to the cavity, as
a function of the detuning $\delta_P=\omega_P-\omega_L$ of the probe frequency
$\omega_P$ from the pump.
For $\delta_c=0$ and $\kappa=0$, the scattering rate of probe photons is
evaluated for a probe Rabi frequency $\tilde{\Omega}_P$ such that
$\tilde{\Omega}_P\ll \bar{g},\Omega,\gamma$. The details of the calculation
are reported in Appendix B. The excitation spectrum is
given by
\begin{equation}
w(\delta_P)=
\gamma \hbar\tilde{\Omega}_P^2\frac{\delta_P^2}
{ [\delta_P(\delta_P+\Delta)-\bar{g}^2]^2+\delta_P^2\gamma^2/4}
\label{Excitation:Rate}
\end{equation}
and it is plotted in Fig.~\ref{Fig:Probe} for two different values of the
detuning $\Delta$ between atom and laser. From Eq.~(\ref{Excitation:Rate}) it is
evident that $w(\delta_P$) vanishes at $\delta_P=0$. This behaviour gives rise
to a Fano--like profile of the excitation spectrum as a function of
$\delta_P$~\cite{Lounis92}, which is visible in Fig.~\ref{Fig:Probe}. This
profile is a manifestation of destructive interference between the excitation
paths contributing to the atomic dynamics, which can be identified with the
absorption of photons from the laser and from the cavity field~\cite{Rice96}.
Interference is at the origin of the two resonances visible in
Fig~\ref{Fig:Probe} where the rate of photon scattering is maximum. They
correspond to values of the probe detuning $\delta_P=\delta_{\pm}$, with
$$\delta_{\pm}=\frac{1}{2}\left(-\Delta\pm\sqrt{\Delta^2+4\bar{g}^2}\right),$$
and have width $\gamma_{\pm}$, which for
$\sqrt{\Delta^2+\bar{g}^2}\gg\gamma/2$ take the simple form
\begin{equation}
\gamma_{\pm}\approx
\frac{\gamma}{4}\left(1\pm\frac{|\Delta|}{\sqrt{\Delta^2+4\bar{g}^2}}\right)
\end{equation}
In the strong coupling regime these resonances correspond to the dressed states
of the atom-cavity system, and their widths determine the characteristic
time--scales of the system's dynamics. Thus, for $\kappa\neq 0$
enhanced cavity emission accompanied by suppression of fluorescence are achieved when
${\rm  min}(\gamma_+,\gamma_-)>\kappa$.

It is remarkable that $w(\delta_P)$ does not depend on $\Omega$ and
thus does
not depend on the average
number of photons inside the cavity. In particular, the
position  and width of the resonances are the ones found for an atom in an
{\it empty} cavity. This result can be simply explained by observing that
the field at the atomic dipole vanishes. In other words, in the reference frame
described by the unitary transformation (\ref{Displace}) the absorption of a
probe photon induces a transition  $|g,0\rangle\to|e,1\rangle$, whereby
$|e,1\rangle$ is the superposition of the eigenstates of the system at
frequencies $\delta_{\pm}$ corresponding to the resonances of the
excitation spectrum. For $\kappa\neq 0$ we have verified that also the
splitting
at the cavity output depends on $g(x)$ and not on $\Omega$. This property
suggests a use of this interference effect in order to probe the atomic
position inside the cavity without significantly perturbing the system.

\subsection{A second atom probing the cavity field}

\begin{figure}[h]
\includegraphics[width=4.5cm]{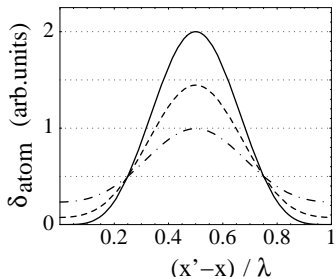}
\caption{a.c.--Stark shift $\delta_{\rm atom}$ of the probing atom
as a function of the distance from
the atom pumping the cavity, which is at an antinode of the standing wave.
Here, $g_0=\Omega=\gamma$, $\Delta=\delta_c=0$, and
the laser propagation direction is perpendicular to the cavity axis.
The detuning of the probe atom
from the cavity frequency is $\Delta_2=1000\gamma$. Solid line: $\kappa=0$;
Dashed line: $\kappa=\gamma$; Dash-Dotted line:
$\kappa=2\gamma$.}
\label{Fig:atom-Probe}
\end{figure}

The electric field inside the cavity can be probed by means of an atom which is
weakly coupled, as shown in Fig.~\ref{Fig:2}(b). This can be, for instance, an
atom of other species whose dipole transition frequency is far--off resonance
from the cavity and the driving laser. This atom thus experiences a small
a.c.--Stark shift $\delta_{\rm atom}$, whose intensity is a function of the
distance $d=x^{\prime}-x$ from the atom which pumps the cavity. It can be
measured with an homodyne detection of the cavity output field, where the pump
field is the local oscillator, or by measuring the fluorescence of the probe
atom. The shift $\delta_{\rm atom}$ is plotted in Fig.~\ref{Fig:atom-Probe} as
a function of the distance $d$ for different values $\kappa$. We observe that
cavity decay tends to cancel the spatial modulation of the total electric
field, which never vanishes inside the cavity for $\kappa\neq 0$. If the
mechanical effects of light are considered, then Fig.~\ref{Fig:atom-Probe}
corresponds to the potential that the probing atom experiences. Hence, the
latter may feel a binding or repulsive force in the vicinity of the first atom,
depending on the sign of the detuning $\Delta_2$ of the laser from the probing
atom resonance.

\section{Scaling with the number of atoms}

\subsection{Two atoms inside the resonator}

So far we have considered that only one atom couples to the cavity mode. If a
second atom of the same kind is inside the cavity and is illuminated by the
laser, the dynamics is in general non-trivial. For $\kappa=0$ suppression of
fluorescence is observed when the atoms are at a distance $\Delta x$ which is
an integer multiple of the wave length $\lambda$. When this occurs, the two
atoms are coupled with the same coupling constant $\bar{g}$ to the cavity mode,
and from Eq.~(\ref{Master:0}) it can be verified that the state
$|g_1,g_2,\beta\rangle$ is the steady state of the system with
$\beta=-\Omega/\bar{g}$. Hence, the total electric field vanishes at both
atoms, while the cavity field is the same as when only one atom couples to the
resonator. In general, one can define the function
\begin{equation}
\label{beta:x}
\beta(x)=\Omega(x)\exp({\rm i}(\pi+kx\cos\theta)) /\bar{g},
\end{equation}
where $\bar{g}=g(x)\neq 0$ and which includes inhomogeneity of the pumping
field. Then, the condition for suppression of fluorescence with two
atoms is fulfilled whenever two positions exist such that
$\beta(x_1)=\beta(x_2)$, where the atoms are located. Clearly, there may
exist parameters regimes for which function (\ref{beta:x}) is not periodic
and a non-trivial solution for suppression of fluorescence with more than one
atom does not exist.

\begin{figure}[h]
\includegraphics[width=8cm]{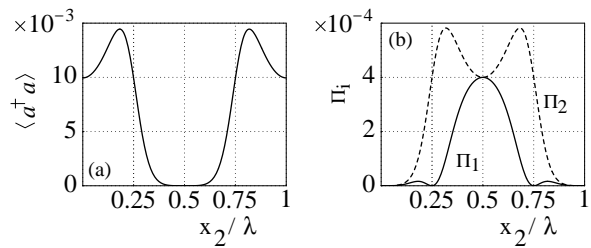}
\caption{(a) mean number of photons and (b) excited state population of atom
  at $x_1$ (solid line) and at $x_2$ (dashed line) as a function of $x_2$.
Here, $x_1$ is an antinode of the standing wave
  $g(x)=g_0\cos kx$. The parameters are $\kappa=0.2\gamma$, $\Omega=\gamma$,
  $g_0=10\gamma$, $\Delta=100\gamma$, $\delta_c=0$, and
the laser propagation direction is perpendicular to the cavity axis.}
\label{Fig:2:atoms:0}
\end{figure}

\begin{figure}[h]
\includegraphics[width=8cm]{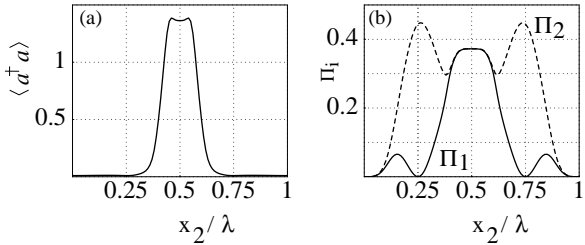}
\caption{Same as Fig.~\protect\ref{Fig:2:atoms:0}. Here, $\kappa=0.01\gamma$, $\Omega=\gamma$, $g=10\gamma$,
  $\Delta=\delta_c=0$, and
the laser propagation direction is perpendicular to the cavity axis. }
\label{Fig:2:atoms:1}
\end{figure}

Figures~\ref{Fig:2:atoms:0} and~\ref{Fig:2:atoms:1} display
the average number of photons and the excited state populations
as a function of the relative distance between the atoms, assuming that
one atom is fixed at the antinode of the standing wave and that
both atoms are homogeneously driven by the laser, which propagates
perpendicularly to the cavity axis, i.e.\ $\theta=\pi/2$. Here it is visible
that the excited state population of both atoms vanishes at $x_2=0,\lambda$.
At these points
the field inside the cavity is different from zero, and it is a local
minimum as a function of $x_2$, as it is particularly evident
in Fig~\ref{Fig:2:atoms:0}. The population of the first atom
vanishes as well when the second atom is at a node of the standing wave.
In this case,
the dynamics of the cavity are determined by the coupling with the atom at
$x_1$, whereas the population of the second atom is determined by the
laser intensity, as if it were in free space.

Note that
Fig.~\ref{Fig:2:atoms:0}
displays the situation when the atoms are driven below saturation. Here
both scatter coherently into the cavity mode and a second type of
interference effect occurs for $x_2=\lambda/2$: At this point the
cavity field vanishes, while the atomic populations are equal and
different from zero. In fact, for $x_2=\lambda/2$ and below saturation
the atoms scatter coherently into the cavity mode with opposite phase.
At saturation, on the other hand,
the scattered light is mostly incoherent, and the cavity
field does not vanish at this point,
as shown in Fig.~\ref{Fig:2:atoms:1}.

\begin{figure}[h]
\includegraphics[width=9cm]{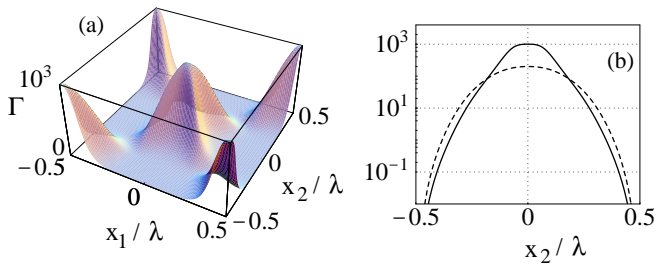}
\caption{(a) Ratio $\Gamma=I_{\rm cav}/I_{\rm at}$ for two atoms as a function
  of their position $x_1$ and $x_2$
inside the cavity for the same parameters as in
Fig.~\protect\ref{Fig:2:atoms:0}. (b) Ratio as in
(a) as a function of $x_2$ for $x_1=0$. The dashed line shows $\Gamma$ for
the same parameters but $\kappa=\gamma$. Note that $x_1,~x_2$ are
plotted modulus $\lambda$, and $x_1\neq x_2$.}
\label{Fig5}
\end{figure}

In Fig.~\ref{Fig5} the ratio between the cavity and fluorescence signal
is displayed for the case illustrated in Fig.~\ref{Fig:2:atoms:0} when
the atoms are driven below saturation and the laser is orthogonal to the
cavity axis. Here one sees clearly that this ratio is maximum when the
atoms are a wavelength apart, namely
where the function~(\ref{beta:x}) assumes the same value
at the atomic positions. For $\kappa\neq 0$
absolute maxima are found when the
atoms are at the antinodes of the cavity mode, where the cooperativity
parameter is largest and the total electric field vanishes.

These considerations can
be extended to three dimensions in a straightforward way. The three-dimensional pattern is found
taking into account the phase of the pump, which we always assume to be
orthogonal to the cavity axis. The zeros of the electric field
are then distributed according to a
Body-Centered-Cubic lattice with distance $\lambda/2$ between adiacent
planes~\cite{Domokos02,Black03}. Fluorescence is suppressed when the atoms are
localized at these points, thus forming a stationary pattern.

\subsection{N atoms inside the resonator}

The dynamics of the coupled system for generic parameters and number of
atoms are
very complex. Nevertheless, insight can be gained in the limit in which
the atoms are driven below saturation. This assumption enables one to
adiabatically eliminate the atomic degrees of freedom from the cavity
equation, and corresponds to the parameter regime $|\gamma/2+{\rm i}\Delta|\gg
\sqrt{N}g,~\sqrt{N}\Omega$, when the collective dipole is driven below
saturation. Under these conditions
the density matrix of the field $\rho_{\rm f}$ obeys the equation
\begin{eqnarray}
  \label{Master:Field} & &\frac{\partial \rho_{\rm f}}{\partial t}
  =\frac{1}{2}(\gamma^{\prime}+\kappa)\left\{2a\rho_{\rm f} a^\dagger-a^\dagger
    a\rho_{\rm f}- \rho_{\rm f}a^\dagger a \right\} \\ & &+{\rm
    i}\delta^{\prime}[a^\dagger a,\rho_{\rm f}] -{\rm i}[(\xi a^\dagger +\xi^*
a),\rho_{\rm f}]\nonumber
\end{eqnarray}
where $\gamma^{\prime}(N)=Ns\gamma$ is the cavity decay rate due to photon
scattering by spontaneous emission, $\delta^{\prime}(N)=\delta_c-Ns\Delta$
contains the a.c.-Stark shift due to the medium, and $\xi$ is the cavity drive
mediated by the dipoles,
\begin{equation}
\xi=Ns\left(\Delta-{\rm
      i}\frac{\gamma}{2}\right) \frac{\sum_n g(x_n)\Omega e^{i\phi_n}}{\sum_n
    |g(x_n)|^2}
\end{equation}
Here, $s=\sum s_n/N$ where $s_n$ is defined for the atom $n$ as
\begin{equation}
  s_n=\frac{g(x_n)^2}{(\gamma/2)^2+\Delta^2}
\end{equation}
From Eq.~(\ref{Master:Field}) it is visible that the system dissipates with
rate $\gamma^{\prime}+\kappa$, which determines the rate at which the steady
state is reached. The steady state
of (\ref{Master:Field}) is $\rho_{\rm f,ss}=|\alpha\rangle\langle\alpha|$,
where $|\alpha\rangle$ is a coherent state with amplitude
\begin{eqnarray}
\alpha &=&-\frac{{\rm i}\xi}{(\gamma^{\prime}(N)+\kappa)/2
+{\rm i}\delta^{\prime}(N)}\nonumber\\
&=&-\Omega \frac{\sum_n g(x_n)
{\rm e}^{{\rm i}\phi_n}}{\sum_n |g(x_n)|^2} \frac{(\gamma/2+{\rm
      i}\Delta)}{(Ns\gamma+\kappa)/2-{\rm i}(\delta_c-Ns\Delta)}
\nonumber\\
\label{beta:max}
\end{eqnarray}
and which is the sum of the electric fields scattered
at each atom. In fact, in this regime
the collective dipole is driven well--below saturation and
radiation is scattered elastically into the cavity mode.

For a large number of
atoms the contributions of each atom sum up so that
the field amplitude $\alpha$ exhibits a narrow peak at the maximum value
$\alpha_0$ as a function of the mean square deviation of
the phase $\Delta\phi_n$. This maximum corresponds to the case when
all atoms scatter in phase, namely when they are distributed at the points
$\{x_1,\ldots,x_N\}$ where the function~(\ref{beta:x}) acquires the same
value. The necessary condition that this situation is verified is that
$\beta(x)$ is periodic, as we have previously observed. In the following we
assume that the laser propagates perpendicularly to the cavity axis, i.e.\
$\theta=\pi/2$. Hence, $\beta(x)$ has periodicity equal to $\lambda$. Assuming
that the atoms scatter in phase into the cavity mode, the amplitude of the
cavity field is given by
\begin{equation}
\alpha_0=-\frac{\Omega}{\bar{g}}
\frac{Ns(\gamma/2+{\rm
    i}\Delta)}{Ns(\gamma/2+{\rm i}\Delta)+\kappa/2-{\rm i}\delta_c}
\label{beta:max:1}
\end{equation}
where $\bar{g}=g(x_1)=\ldots=g(x_N)$ and $\bar{g}\neq 0$,
namely the atoms are spatially distributed in a pattern which
has spatial periodicity $\lambda$. The pattern is here assumed to have low
filling factor, such that sub- and supperradiance effects in the scattering in
free space are negligible. Note that for large filling factors
subradiance may give rise to other meta-stable states of the collective
dynamics.

\subsubsection{Stability of the atomic patterns}

The atomic pattern in a standing wave cavity is invariant per translation by $\lambda$.
In principle, there is an infinite number of patterns for any value of $x$ in the range
$[0,\lambda)$, corresponding to different values of the coupling constant $g(x)$.
When the mechanical effects of light are taken into account, however, the
equilibrium positions are at the antinodes of the cavity-mode standing wave
$x=0$ or $x=\lambda/2$. At these
points, in fact, the force on the atoms vanishes. This is evident when we
consider the force $F_n$ entering the semiclassical equation of the motion for
the atom at $x_n$ moving along the cavity
axis~\cite{Domokos01,Domokos02,RitschReview}
\begin{eqnarray}
\dot{F}_n   &=&\hbar k U_0 |\alpha|^2 \sin (2kx_n)\nonumber\\
            & &+2\hbar k {\rm
  Im}\{\eta_{\rm eff}^*\alpha\}\sin (kx_n)
\label{eq:p}
\end{eqnarray}
where $U_0=g_0^2\Delta/(\Delta^2+\gamma^2/4)$ is the light shift due to the
coupling to the cavity,
$\Gamma_0=g_0^2\gamma/2(\Delta^2+\gamma^2/4)$ the rate of dissipation,
and $\eta_{\rm eff}=\Omega g_0/(-{\rm i}\Delta+\gamma/2)$ the term due to the
transversal pump on the atoms. The parameter $\alpha$
describes the field amplitude, which evolves according
to Eq.~(\ref{Master:Field}). We denote with even (odd) pattern the atomic pattern
where the atoms are localized at the equilibrium points $x_n^{(0)}$
such that $\cos kx_n^{(0)}=1$ ($\cos kx_n^{(0)}=-1$). For $\kappa=0$ and
$\delta_c=0$, one has suppression of fluorescence when the cavity field
coherent state has amplitude $\alpha=\beta_{\ell}=\Omega/g_0 {\rm e}^{{\rm
    i}(\ell+1)\pi}$, with $\ell=0,1$ depending on whether the pattern is even
or odd. Hence, the cavity fields due to each pattern differ
by a phase $\pi$. Numerical studies
have reported selforganization of the atoms in these patterns~\cite{Domokos02}.
Stability is found
for small (but non-vanishing) negative values of $\delta_c$, as we have
verified numerically~\cite{Zippilli04}. From Eq.~(\ref{eq:p}) we can
estimate the force $\delta f_n$ around these points when the atoms undergoes a
small displacement $\delta x_n$ from the equilibrium
position $x_n^{(0)}$. At first order in $\delta x_n$ the force takes the form
\begin{equation}
\label{Fluctua}
\delta f_n\sim 2\hbar k^2\left(\frac{\Omega}{g_0}\right)^2\frac{\delta_c}{N}
\delta x_n
\end{equation}
and it is clearly a restoring force for $\delta_c<0$. This condition is
sufficient, since the field amplitude $\alpha$ does not vary in first order in
$\delta x_n$, nor does the force for small fluctuations in $\alpha$.
Remarkably, Eq.~(\ref{Fluctua})
is independent of $\Delta$. Moreover, its intensity
depends on the mean number of cavity photons, and thus on the pump intensity.
This result has been confirmed by numerical simulations and
is in line with the experimentally observed dependence of
enhanced cavity emission on the intensity of the pump, showing that the effect
manifests itself
when the pump intensity exceeds a threshold value~\cite{Chan03}.
We remark that result~(\ref{Fluctua}) is valid when the ratio $\delta_c/N$ is
sufficiently small, so that to good approximation the field inside the cavity
is given by $\beta_{\ell}=(-1)^{\ell +1}\Omega/g_0$ and the total field at the
atomic positions almost vanishes. The dependence of the system dynamics on the
atom number $N$ is discussed in the following subsection.

\subsubsection{The cavity field when the atoms emit in phase}

\begin{figure}[h]
\includegraphics[width=9cm]{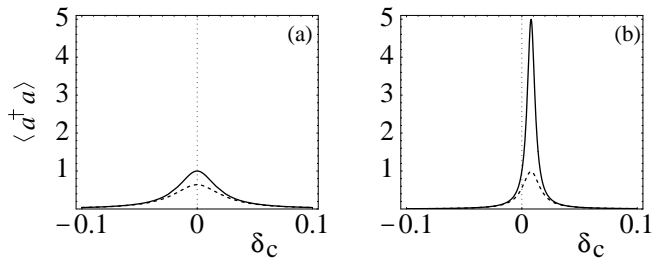}
\caption{Mean number of cavity photons as a function of $\delta_c$ in units of
  $\gamma$. The solid line corresponds to
$\kappa=0$, the dotted line to $\kappa=0.01\gamma$.
For $N=1$ atom, the parameters are
$\Omega=g=0.1\gamma$, and (a) $\Delta=0$, (b) $\Delta=\gamma$.}
\label{Fig:alpha}
\end{figure}

We now assume that the atoms are localized in an even pattern, such that the
cavity field amplitude is
$\beta_{\ell}=\beta_0=-\Omega/g_0$, and disregard the
mechanical effects of light. For $\kappa=0$ and
$\delta_c=0$ we recover from Eq.~(\ref{beta:max:1}) the result
$\alpha_0=\beta_0$. Thus, in this limit the stationary field is
independent of the number $N$ of atoms and of the detuning $\Delta$
between laser and dipole transition.
The field amplitude achieves the maximum value as a function of $\delta_c$
for $\delta^{\prime}=0$, corresponding to the
condition $\delta_c=Ns\Delta$.  This is visible in Fig.~\ref{Fig:alpha}, where
the average number of photons is plotted as a
function of $\delta_c$ and for two different values of $\Delta$.
For $\Delta\neq 0$ the detuning $\delta_c=Ns\Delta$
is the a.c.-Stark shift of the cavity mode frequency
due to the coupling with the atomic dipoles. For this value the
classical field drives the system resonantly, and the amount of energy
transferred into the cavity mode is maximum. Note that the position of the
resonance $\delta_c=Ns\Delta$
scales linearly with $N$ and for $|\Delta|\gg \gamma$ is inversely
proportional to $\Delta$. The corresponding linewidth
$\gamma^{\prime}+\kappa$ scales as $N/\Delta^2$ for
$\gamma^{\prime}\gg\kappa$. Obviously, large values of $\kappa$
broaden the resonances.

\begin{figure}[h]
\includegraphics[width=9cm]{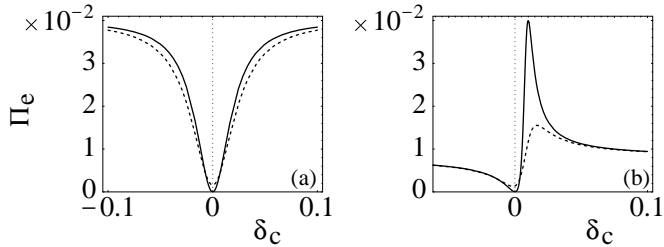}
\caption{Excited state population as a function of $\delta_c$
in units of $\gamma$,
Same parameters as in Fig.~\ref{Fig:alpha}.}
\label{Fig:Pe}
\end{figure}

At the amplitude of the cavity field of
Eq.~(\ref{beta:max:1}) the population of the excited state of an atom in any
of the pattern positions is given by
\begin{equation} \label{Excited}
\Pi_{e}=
  \frac{\Omega^2}{(\gamma/2)^2+\Delta^2}
  \frac{\kappa^2/4+\delta_c^2}{(\gamma^{\prime}+\kappa)^2/4+\delta^{\prime 2}}
\end{equation}
and is displayed in Fig.~\ref{Fig:Pe} as a function of
$\delta_c$ for some parameter
regimes. Clearly, for $\kappa=0$ and $\delta_c=0$ the
excited state population vanishes, indicating that the atoms stop fluorescing.
For $\Delta\neq 0$ the population $\Pi_e$ exhibits a maximum, which is located
at $\delta_c\sim Ns\Delta (1+\gamma^2/4\Delta^2)$ for $|\Delta|\gg \gamma$.
For $\kappa\neq 0$
the center-frequencies of the resonances are shifted by an amount
proportional to the cavity decay rate, the
curves are broadened, and the excited states population does not vanish
at $\delta_c=0$.

The behaviour of the system as the number of atoms
$N$ is varied exhibits remarkable features. In fact, $N$ appears in the
denominator of Eqs.~(\ref{beta:max:1}) and~(\ref{Excited}), scaling the atomic
effects in the cavity dynamics. In particular, a critical value $N_0$ for the
number of atoms can be
identified, such that for $N\gg N_0$ the coupling with the atoms affects
relevantly the cavity dynamics, whereas for $N\ll N_0$ atoms and cavity are
weakly coupled.

\begin{figure}[h]
\includegraphics[width=4.5cm]{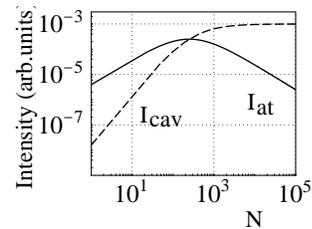}
\caption{$I_{\rm cav}$ and total $I_{\rm at}$ as a
 function of $N$ for
$\Omega=g=\kappa=10^{-3}\gamma$, and $\Delta=\delta_c=0$.}
\label{N:dependence:1}
\end{figure}

For $\Delta=\delta_c=0$ one finds the value
$N_0=\kappa/s\gamma=1/2C_1$ where $C_1=2g_0^2/\kappa\gamma$ is the one-atom
cooperativity parameter~\cite{Kimble94}. Thus for $N\gg N_0$ the system is
characterized by a large cooperativity parameter. In particular, when
$N\ll N_0$ the excited states population in Eq.~(\ref{Excited}) acquires
approximately  the value as in
free space, while the cavity field amplitude scales linearly with the number
of atoms. There is thus no back--action of the cavity on the atomic dynamics,
since the cavity decay rate is faster than the rate at which the atomic
degrees of freedom reach their steady state. On the other hand, when $N\gg
N_0$ the field amplitude tends to the asymptotic value $\alpha\to-\Omega/g$,
while $\Pi_{e}\bigl|_0\propto \kappa^2/N^2$. Thus, the power dissipated by
spontaneous emission scales with $1/N$, while the signal at the cavity output
is constant and independent on the number of atoms.
Figure~\ref{N:dependence:1} displays
the signal at the cavity output and the total fluorescence
signal evaluated from Eqs.~(\ref{beta:max:1}) and~(\ref{Excited}), respectively
as a function for $N$. For these parameters $N_0\sim 10^3$.

\begin{figure}[h]
\includegraphics[width=9cm]{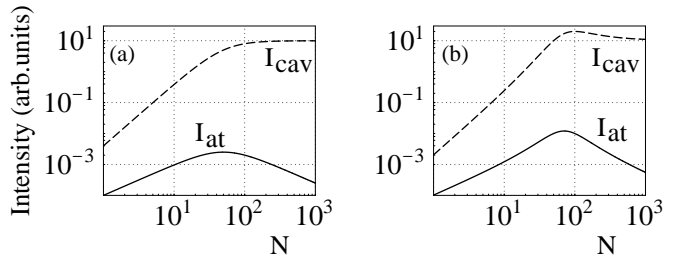}
\caption{$I_{\rm cav}$ and total $I_{\rm at}$ as a
 function of $N$ for
$\Omega=g=\kappa=10\gamma$, $\Delta=-1000\gamma$ and (a)
$\delta_c=0$, (b) $\delta_c=-5\gamma$}
\label{N:dependence:2}
\end{figure}

An analogous behaviour can be found for large values of $\Delta$, and is
illustrated in Fig.~\ref{N:dependence:2}. For $|\Delta|\gg \gamma,\kappa$ the
critical number of atoms, determining the regime of strong coupling, is given
by $N_{0,\Delta}=|\Delta|\kappa/g^2$, and enhanced cavity emission accompanied
by suppression of fluorescence is observed for $N\gg N_{0,\Delta}$. This
behaviour is also found for values of the detuning $\delta_c\neq 0$, as shown
in Fig.~\ref{N:dependence:2}(b) for $\delta_c=-5\gamma$, provided that $N$ is
sufficiently large to fulfill the relation $|\delta_c|\ll Ns|\Delta|$. Note
that for values of $N$ such that $\delta_c=Ns\Delta$, namely when a collective
state of the system is driven resonantly, the signals in
Fig.~\ref{N:dependence:2}(b) exhibit a maximum. Nevertheless, as $N$
increases, $I_{\rm cav}$ tends asymptotically to the value $I_{\rm cav}\to
\kappa|\Omega/g_0|^2$ which is independent, among others, of $N$ and of
$\Delta$. It should be noted that in this case increasing $N$ corresponds to
increasing the width of the window around the value $\delta_c=0$ appearing in
the atomic population as a function of $\delta_c$, as shown in
Fig.~\ref{Fig:Pe}. Thus, the condition $|\delta_c|\ll Ns|\Delta|$ corresponds
to values of the detuning $\delta_c$ which are much smaller than the
a.c.-Stark shift, hence for which the condition of destructive intereference
is still (although approximately) fulfilled.

The parameter regimes discussed in Fig.~\ref{N:dependence:2}(b) are consistent
with the ones of the experiment by~\cite{Black03}, that reported a
rate of emission into the cavity modes exceeding by orders of magnitude the
rate of fluorescence into the modes external to the cavity. This observation
was accompanied by the measurement of a coherent cavity field whose
characteristic gave evidence of atomic self-organization. This behaviour has
been explained as Bragg scattering of the pump light by the atomic
grating. However, from the results presented in this section we can argue
that Bragg scattering is actually suppressed in this regime, as the
experimental regime of~\cite{Black03} can be classified to be in the region with $N\gg
N_0$. In fact, in the strong coupling regime a cavity field establishes when the atoms
organize spatially, that cancels out with the pump at the atomic positions. As a consequence
the atoms decouple from the cavity and pump field, and are in the ground state. Therefore,
there is no fluorescence nor superradiant scattering into the cavity mode. In this regime
the main source of dissipation is through
cavity decay. We remark that these dynamics is encountered also in the case of a
single atom. In fact for strong coupling the stationary cavity field is solely
determined by pump intensity and cavity coupling at the atomic position, while
the atoms are in the ground state. In particular, the regime of large $N$ in
Figs.~\ref{N:dependence:1}-\ref{N:dependence:2} corresponds to large cooperativity
parameters, while in Fig.~\ref{Fig:Compare:2}
strong coupling is achieved for small
$\kappa$. Note that, differently from optical
bistability~\cite{OpticalBistability}, where bistable dynamics are observed
for large cooperativity, here there is only one steady state,
where the atoms are in the ground state.

In summary, the enhanced cavity emission of~\cite{Black03} can be
traced back to an interference effect between pump and cavity field, which
is established
for large cooperativity parameters. Although in this treatment we
have neglected the center-of-mass motion, this hypothesis is supported by the
stability of the pattern in this parameter regime.

Finally, it should be observed that one important condition for these dynamics
is that the atoms are localized according to a stationary pattern. This
condition constitutes a substantial difference to the collective scattering
via acceleration observed in the dynamics of the collective atomic recoil
laser~\cite{Zimmerman03,CARL}.

\section{Discussion and outlook}

Collective effects play an important role in the dynamics of N coherently
driven atoms within a cavity mode. For a resonant laser and a sufficiently
large cooperativity parameter the atomic scattering of photons into the cavity
field may exceed the scattering into the free-space modes by several orders of
magnitude despite weak coupling and low excitation of each individual atom. In
this regime the cavity output exhibits Poissonian photon statistics
independent of the mean cavity photon number. In addition the probe excitation
spectra of the atoms reveal the cavity vacuum-Rabi splitting even for strong
pump fields, when the mean number of cavity photons is large.

The phenomenon can be understood as interference between the transverse pump
and the cavity field, which is established when the coupling between atoms and
cavity mode is sufficiently stronger than other effects determining the
system's dynamics. Remarkably, the establishing of this regime corresponds to
the situation in which the total electric field at the atomic position
vanishes. For two or more atoms inside the cavity these conditions are
accessed when the atoms are distributed  according to a spatial pattern with
periodicity equal to the mode wavelength. By means of a simple model we have
shown that stable patterns are achieved for suitable laser and cavity
parameters, when the locations of the pattern are at the antinodes of the
cavity standing wave. We have identified two patterns, which correspond to
fields inside the cavity which are shifted by a phase $\pi$. The results
predicted by this model are in qualitative agreement with the dynamics
reported in~\cite{Domokos02,Chan03,Black03}, and provide a physical picture of
the phenomena observed. It should be noted that the cavity used
in~\cite{Chan03,Black03} is multimode, whereas in this work we consider a
single mode cavity. Nevertheless the dynamics reported in~\cite{Black03} can
be reproduced with a model consisting in a single mode cavity and two level
atoms, showing that the basic physical phenomena can be traced back to the
interference effect discussed here.

The phenomenon of interference in the driven Jaynes-Cummings model has been
denoted with "cavity induced transparency" by Rice and Brecha~\cite{Rice96},
and it can be traced back to the classical dynamics of two coupled damped
oscillators~\cite{Nussenzveig01}. Rice and collaborators have theoretically
investigated the dynamics of a classical dipole coupling resonantly to a
cavity mode when the cavity is driven~\cite{Brecha99,Clemens00}. In this case
the field due to atomic polarization cancels out with the drive on the
cavity. Due to this effect the cavity electric field vanishes. Thus, the
phenomenon is established when the dipole decay rate is smaller than the
cavity decay rate and, once this regime is accessed, energy is dissipated mainly
by spontaneous decay. This situation might seem equivalent in many respects to
the case discussed by Carmichael and coworkers in~\cite{Alsing92}, where the
role of cavity and atom are exchanged. Nevertheless, when the cavity is driven
quantum noise and saturation effects on the dipole give rise to deviations
from its classical behaviour and thus from interference. Here interference is
recovered for a sufficiently large number of dipoles $N$, so that the
collective dipole is to good approximation an oscillator. Another interesting
difference between the driven-cavity and the driven-atom case is the signal at
the cavity output. When the cavity is driven and $\gamma\ll\kappa$, the
$g^{(2)}(\tau)$ function is antibunched at $\tau=0$~\cite{Brecha99,Clemens00}. On the
contrary, when the atoms are driven and $\kappa\ll\gamma$, we have shown that
$g^{(2)}(0)=1$ even when the mean energy of the cavity field energy is very
small.

It is instructive to compare the phenomenon of suppression of fluorescence
investigated in this work with the phenomenon of electromagnetically induced
transparency manifesting itself in driven multilevel atomic
transitions~\cite{EIT}. The two types of interference arise because of
different dynamics: In EIT the atomic polarization is orthogonal to the field
polarization, so that the atom does not absorb photons. In "cavity-induced
transparency" the laser and the cavity field cancel out, so that the total
electric field at the atom is zero. It is this very property that leads to the
vacuum Rabi splitting observed in the excitation spectrum even when the mean
energy of the cavity field is significantly large.

There are several interesting questions which are worth investigating starting
from these results. For instance, do other patterns exist, than the ones
found, which may be stable and bring to a different steady state of the cavity
field? In this context,we have considered a transverse pump whose propagation
direction is orthogonal to the cavity axis. In this way, the periodicity of
the pattern is determined solely by the periodicity of the cavity standing
wave. When the angle between cavity and pump is different the situation may
change drastically, even giving no a priori possibility of finding a stable
pattern for more than one atom. For other cases, when the cavity mode is not a
standing wave, but, say, a ring cavity~\cite{Hemmerich03,Zimmerman03}, again
other dynamics are expected. Such questions will be tackled by treating
systematically the mechanical effects of light-atom interaction, and will be
subject of following works~\cite{Zippilli04}.

For systems of one or few atoms in high-Q cavities the interference phenomenon
presents several potentialities for implementing coherent dynamics of quantum
systems. For instance, the vacuum Rabi splitting observed by probing the
system depends on the position of the atom in the mode, and may allow to
determine the spatial mode structure~\cite{MPQ01}, as well as to implement
feedback schemes on the atomic motion~\cite{FeedbackRempe,Mancini}. Moreover,
several experimental setups can presently trap single or few atoms and couple
them in a controlled way to the cavity
field~\cite{Eschner01,MPQ01,Mundt02,Sauer03,KimbleFORT03}. The dynamics
discussed here can be applied for instance to implementations of quantum
information processing, since interference effects are rather robust against
noise and decoherence. In addition, the coherence properties of the
transmitted signal, which are preserved even for very small photon numbers,
suggest an alternative kind of photon-emitters to the one investigated
in~\cite{Kuhn02,Kimble03a,Kimble03b}.

\begin{acknowledgements}
The authors gratefully ackowledge discussions with
H.J.\ Carmichael, P.\ Domokos, A.\ Kuhn, H.\ Mabuchi, P.\ Pinkse, G.\ Rempe,
and W.P.\ Schleich. This work has been supported by the TMR-network QUEST, the
IST-network QGATES and the Austrian FWF project S1512.
\end{acknowledgements}

\begin{appendix}
\section{Evaluation of the perturbative corrections in $\kappa$}

In this appendix we present the main steps to determine $\rho(t)$ in
(\ref{Sol:1}) to second order in $\kappa$. In the
displaced frame defined by the unitary transformation (\ref{Displace})
$\tilde{\rho}(t)={\cal D}(\beta)\da\rho(t){\cal D}(\beta)$, and
Eq.~(\ref{Sol:1}) takes the form
\begin{eqnarray*}
\tilde{\rho}(t)=\tilde{\cal S}(t)\tilde{\rho}(0)
  +\int_0^t{\rm d}\tau \tilde{\cal S}(t-\tau)(\kappa\tilde{\cal  K}_0+\gamma
  J) \tilde\rho(\tau)
\end{eqnarray*}
where $\tilde{\rho}(0)={\cal
  D}(\beta)\da\rho_{\rm ss}{\cal D}(\beta)=|g,0\rangle\langle g,0|$ and
\begin{eqnarray*}
&&\tilde{\cal  K}_0X=aXa\da-\frac{\Omega}{g}(aX+X
  a\da) +\left(\frac{\Omega}{g}\right)^2X\\
&&\tilde{S}(t)X=U_{\rm eff}(t)X
U_{\rm eff}(t)\da \\
&& U_{\rm eff}(t)=\exp\left(-\frac{1}{{\rm i}\hbar}
\tilde{H}_{\rm eff}t\right)
\end{eqnarray*}
Here, $\tilde{\cal
  K}_0$, $\tilde{S}(t)$ are the transformed superoperators defined on a
density matrix $X$ and $U_{\rm eff}(t)$ is a non--unitary operator, with
$\tilde{H}_{\rm eff} ={\cal D}(\beta)\da H_{\rm eff}{\cal D}(\beta)$. The
operator $\tilde{H}_{\rm eff}$ is non--Hermitian. Its spectrum and
eigenvectors are calculated
by solving the secular equations for the right and left
eigenvectors of $\tilde{H}_{\rm eff}$, according to $\tilde{H}_{\rm
  eff}|v_{\lambda}\rangle=\hbar\lambda|v_{\lambda}\rangle$,  $\tilde{H}_{\rm
  eff}^{\dagger}\overline{|v_{\lambda}\rangle}=\hbar\lambda^*
\overline{|v_{\lambda}\rangle}$.
  The states $\{|v_{\lambda}\rangle,\overline{\ke{v_{\lambda}}}\}$  constitute
  a biorthogonal basis, such that
  $\sum_{\lambda}|v_{\lambda}\rangle\overline{\langle v_{\lambda}|}=1$. In
  this basis the operator $U_{\rm eff}(t)$ can be written as
\begin{equation}
U_{\rm eff}(t)=\sum_{\lambda} {\rm e}^{-{\rm i}\lambda
  t}|v_{\lambda}\rangle\overline{\langle v_{\lambda}|}
\end{equation}
We expand now $U_{\rm eff}(t)$ in second order in the parameter $\kappa$,
$U_{\rm eff}(t)=U_{\rm eff}^{(0)}(t)+\kappa U^{(1)}_{\rm
  eff}(t)+\kappa^2U^{(2)}_{\rm eff}(t)$, where the superscript indicates
the corresponding order in the perturbative expansion. In order to evaluate
these terms we define
\begin{equation}
\label{H:eff:tilde}
\tilde{H}_{\rm eff} =\tilde{H}_{\rm
  eff}^{(0)}+\kappa V
\end{equation}
with
\begin{eqnarray}
\label{H:eff:0}
\tilde{H}_{\rm eff}^{(0)} &=&\hbar g(a\sigma\da
+a\da\sigma)-\hbar\left(\Delta
+{\rm  i}\frac{\gamma}{2}\right)|e\rangle\langle e|
\\ V &=&-\frac{{\rm i}\hbar}{2}\pt{ a\da a+\frac{\Omega^2}{g^2}}+ \frac{{\rm
    i}\hbar}{2}\frac{\Omega}{g}(a\da +a)
\end{eqnarray}
and solve the eigenvalue equation at second order in $\kappa$, obtaining
\begin{eqnarray*}
&&\lambda=\lambda^{(0)}+\kappa \lambda^{(1)}+\kappa^2
  \lambda^{(2)}+{\rm o}(\kappa^3)
\\ &&|v_{\lambda}\rangle={\cal N}^{-1/2}(|v_{\lambda}^{(0)}\rangle+\kappa
  |v_{\lambda}^{(1)}\rangle  +
  \kappa^2|v_{\lambda}^{(2)}\rangle+{\rm o}(\kappa^3))
\end{eqnarray*}
and analogously for the
left eigenvectors, where ${\cal N}=\overline{\langle
  v_{\lambda}}|v_{\lambda}\rangle$.
In particular, the solutions at zero order have the form
\begin{equation}
\lambda_{n,\pm}^{(0)}=-\frac{1}{2}\pt{\Delta+{\rm
      i}\frac{\gamma}{2} \mp\sqrt{\left(\Delta+{\rm
          i}\frac{\gamma}{2}\right)^2+4g^2n}}
\end{equation}
with the respective right eigenvectors
\begin{equation}
\ke{v_{n\pm}^{(0)}}=a_{n,\pm}\ke{e,n-1}+b_{n,\pm}\ke{g,n}
\end{equation}
and
\begin{eqnarray*}
&&a_{n,\pm}=\lambda_{n,\pm}{\cal N}_{n,\pm}^{-1/2}\\
&&b_{n,\pm}=g\sqrt{n} {\cal N}_{n,\pm}^{-1/2}
\end{eqnarray*}
while the left eigenvectors $\overline{\ke{v}}$ have the form
$\overline{\ke{v_{n,\pm}}}=a_{n,\pm}^*\ke{e,n-1}+b_{n,\pm}^*\ke{g,n}$, where
${\cal N}_{n,\pm}=\overline{\langle v_{n,\pm}}|v_{n,\pm}\rangle$. Using this
expansion, the terms of the perturbative expansion of the operator $U_{\rm
  eff}(t)$ are immediately found, and the evaluation of
$\tilde{\rho}(t)=\tilde{\rho}^{(0)}(t)+\kappa\tilde{\rho}^{(1)}(t)+
\kappa^2\tilde{\rho}^{(2)}(t)+{\rm o}(\kappa^3)$, consists in the evaluation
of the integrals
\begin{eqnarray}
\label{rho(t)}
&& \tilde{\rho}^{(0)}(t)={\cal S}^{(0)}(t)|g,0\rangle\langle
  g,0| \\
&& \tilde{\rho}^{(1)}(t)=\int_0^t{\rm d}\tau {\cal
    S}^{(0)}(t-\tau){\cal K}_0 {\cal S}^{(0)}(\tau)|g,0\rangle\langle g,0|
\nonumber\\
  && \tilde{\rho}^{(2)}t)={\cal S}^{(2)}(t)|g,0\rangle\langle
  g,0|+\int_0^t{\rm d}\tau \Bigl[{\cal S}^{(0)}(t-\tau) J {\cal S}^{(2)}(\tau)
\nonumber
\\ && + {\cal
      S}^{(0)}(t-\tau){\cal K}_0 {\cal S}^{(1)}(\tau)+{\cal
      S}^{(1)}(t-\tau){\cal K}_0 {\cal S}^{(0)}(\tau)
  \Bigr]|g,0\rangle\langle g,0| \nonumber
\end{eqnarray}
where ${\cal S}^{(l)}(t)X=\sum_{p=0}^lU_{\rm eff}^{(l-p)}(t)XU^{(p)}_{\rm
  eff}(t)^{\dagger}$. Note that in (\ref{rho(t)}) we have
  omitted to write the terms containing $J{\cal
  S}^{(j)}(\tau)|g,0\rangle\langle g,0|$ (with $j=0,1$), since they vanish.

\section{Evaluation of the excitation spectrum}

We evaluate the excitation spectrum by calculating the transition amplitude
which describes the scattering of a probe photon
into the modes of the electromagnetic field, into which the dipole
spontaneously emit. The modes of the electromagnetic field are here treated
quantum mechanically. The Hamiltonian determining the dynamics is
\begin{equation}
H^{\prime}=H+H_{\rm probe}+H_{\rm emf}
\end{equation}
where $H$ is defined in Eq.~(\ref{H}), $H_{\rm
  probe}=\hbar\delta_Pb^{\dagger}b + V$, with $b$, $b^{\dagger}$ annihilation
  and creation operators of a probe photon, $\delta_P$ detuning of the probe
  from the cavity frequency, and
\begin{equation}
V=\hbar\Omega_P\left(b\sigma_n^{\dagger}+b^{\dagger}\sigma_n\right)
\end{equation}
the interaction of the probe with the dipole, with $\Omega_P$ vacuum Rabi
frequency.
The term $H_{\rm emf}$ describes
the coupling of the dipole to the other external modes of the electromagnetic
field , $H_{\rm emf}=\sum_{\bf k}\hbar\delta_{\bf k}b_{\bf
k}^{\dagger}b_{\bf k}+W$, where ${\bf k}$
labels the mode at frequency $\delta_{\bf k}$ (in the reference frame of the
drive) and wave vector ${\bf k}$, with corresponding creation and
annihilation operators $b_{\bf k}^{\dagger}$, $b_{\bf k}$, and
$W$ describes the interaction with the atomic dipole,
\begin{equation}
W=\sum_{\bf k}\hbar g_{\bf k}\left(\sigma^{\dagger}b_{\bf k}+ \sigma b_{\bf
    k}^{\dagger}\right)
\end{equation}
Here, $g_{\rm k}$ is the vacuum Rabi frequency for the coupling of the mode to
the dipole. \\

We assume that at $t=0$ the system is in the stationary state of the driven
dipole and cavity system, the probe field is a coherent state of amplitude
$\eta$ such that $|\eta|^2\ll 1$ and the other modes of the
electromagnetic field are in the vacuum $|0_{\bf k}\rangle$.
As we are interested in the probability that a probe photon is
scattered into the modes of the e.m.f.-field, the initial state
is given with probability $|\eta|^2$ by
\begin{equation}
|\psi_i\rangle=|g,\beta;1_P,0_{\bf k}\rangle
\end{equation}
and it is at energy $E_i=\hbar\Delta+\hbar\delta_P$, while the final state is
\begin{equation}
|\psi_{f,{\bf k}}\rangle=|g,\beta;0_P,1_{\bf k}\rangle
\end{equation}
The transition amplitude is the element of the scattering matrix $S_{i,f_{\bf
    k}}$,
\begin{equation}
\label{Sfi}
S_{if_{\bf k}}
=-2{\rm i}\pi\lim_{T\rightarrow\infty}\delta^{(T)}(E_{f_{\bf
    k}}-E_i)T_{if_{\bf k}}(E_i)
\end{equation}
where $\delta^{(T)}(E)$ is the diffraction function,
\begin{equation}
\delta^{(T)}(E)=\frac{1}{\pi}\frac{\sin (ET/2\hbar)}{E}
\end{equation}
and $T_{fi}(E_i)$ is the transition matrix element, which at lowest
non-vanishing order has the form
\begin{equation}
\label{T:fi}
T_{if_{\bf k}}(E_i)=\langle \psi_{f,{\bf k}}|W\frac{1}{E_i-H_{\rm eff}}
V|\psi_i\rangle
\end{equation}
Here, $H_{\rm eff}^{\prime}$ is the effective Hamiltonian,
\begin{equation}
H_{\rm eff}^{\prime}=H_{\rm eff}+\hbar\delta_Pb^{\dagger}b
\end{equation}
where $H_{\rm eff}$ is given in~(\ref{H:eff}).
At lowest order in $\eta$, the transition matrix element (\ref{T:fi}) has the
form
\begin{eqnarray}
T_{if_{\bf k}}(E_i)
=\hbar^2g_{\bf k}\tilde{\Omega}_P\langle e,\beta;0_P,0_{\bf k}
|\frac{1}{E_i-H_{\rm eff}^{\prime}}|e,\beta;0_P,0_{\bf k}\rangle\nonumber\\
\label{T:fi:2}
\end{eqnarray}
where $\tilde{\Omega}_P=\eta\Omega_P$.
The solution of (\ref{T:fi:2}) can be easily
found in the reference frame defined by (\ref{Displace}): Here, $H_{\rm
  eff}^{\prime}=\tilde{H}_{\rm eff}^{(0)}+\hbar\delta_Pb^{\dagger}b$, where we
have used (\ref{H:eff:0}). Finally, we obtain
\begin{eqnarray}
T_{if_{\bf k}}(E_i)
=\hbar g_{\bf k}\tilde{\Omega}_P\frac{\delta_P}
{\delta_P[\delta_P+\Delta+{\rm i}\gamma/2]-g^2}
\label{T:fi:3}
\end{eqnarray}
By substituting this result in Eq.~(\ref{Sfi}) we find the transition
amplitude. The rate~(\ref{Excitation:Rate}) is found from (\ref{Sfi})
after summing over all modes of the continuum and taking the modulus squared
divided by the time $T$~\cite{Lounis92}.

\end{appendix}


\begin{thebibliography}{99}

\bibitem{Pinkse04}
P.\ Maunz, T.\ Puppe, I.\ Schuster, N.\ Syassen, P.W.H.\ Pinkse and G.~Rempe, Nature {\bf 428}, 50, 2004

\bibitem{Hood00}
C.J.\ Hood, T.W.\ Lynn, A.C.\ Doherty, and H.J.\ Kimble, Science {\bf
  287}, 1447 (2000).

\bibitem{Pinkse00}
P.W.H.\ Pinkse, T.\ Fisher, P.\ Maunz, and G.\ Rempe, Nature (London) {\bf
  404}, 365 (2000).

\bibitem{Mundt02}
A.B.\ Mundt, A.\ Kreuter, C.\ Becher, D.\ Leibfried, J.\ Eschner, F.\
Schmidt-Kaler, and R.\ Blatt, Phys.\ Rev.\ Lett.\ {\bf 89}, 103001 (2002).

\bibitem{Sauer03}
J.A.\ Sauer, K.M.\ Fortier, M.S.\ Chang, C.D.\ Hamley, M.S.\ Chapman, Phys.\ Rev.\ A
{\bf 69} 051804 (2004).

\bibitem{KimbleFORT03}
J.\ McKeever, J.R.\ Buck, A.D.\ Boozer, A.\ Kuzmich, H.C.\ N\"agerl, D.M.\
Stamper-Kurn, H.J.\ Kimble, Phys.\ Rev.\ Lett.\ {\bf 90}, 133602 (2003).


\bibitem{Kimble95}
Q.A.\ Turchette, C.J.\ Hood, W.\ Lange, H.\ Mabuchi, H.J.\ Kimble, Phys.\
Rev.\ Lett.\ {\bf 75}, 4710 (1995).

\bibitem{Kuhn02}
A.\ Kuhn, M.\ Hennrich, and G.\ Rempe, Phys.\ Rev.\ Lett.\ {\bf 89}, 067901
(2002).

\bibitem{Kimble03a}
A.\ Kuzmich, W.P.\ Bowen, A.D.\ Boozer, A.\ Boca, C.W.\ Chou,
L.-M.\ Duan, H.J.\ Kimble, Nature (London) {\bf 423}, 731 (2003).


\bibitem{Kimble03b}
J.\ McKeever, A.\ Boca, A.D.\ Boozer, J.R.\ Buck, H.J.\ Kimble, Nature
(London) {\bf 425}, 268 (2003).


\bibitem{MPQ01}
G.R.\ Guth\"ohrlein, M.\ Keller, K.\ Hayasaka, W.\ Lange, and H.\ Walther,
Nature {\bf 414}, 49 (2001).

\bibitem{Horak02}
P.\ Horak, H.\ Ritsch, T.\ Fischer, P.\ Maunz, T.\ Puppe, P.W.H.\ Pinkse, G.\
Rempe, Phys.\ Rev.\ Lett.\ {\bf  88}, 043601 (2002).


\bibitem{Zimmerman03}
D.\ Kruse, C.\ von Cube, C.\ Zimmerman, Ph.W.\ Courteille, Phys.\ Rev.\ Lett.\
{\bf 91}, 183601 (2003).

\bibitem{Hemmerich03}
B.\ Nagorny, Th.\ Els\"asser, A.\ Hemmerich, Phys.\ Rev.\ Lett.\ {\bf 91},
153003 (2003).

\bibitem{Chan03}
H.W.\ Chan, A.T.\ Black, V.\ Vuletic, Phys.\ Rev.\ Lett.\ {\bf 90}, 063003
(2003).

\bibitem{Black03}
A.T.\ Black, H.W.\ Chan, V.\ Vuletic, Phys.\ Rev.\ Lett.\ {\bf 91}, 203001
(2003).

\bibitem{Zippilli04a}
S.\ Zippilli, G.\ Morigi, H.~Ritsch, Phys. Rev. Lett. {\bf 93},
123002 (2004).

\bibitem{Eschner01}
J.\ Eschner, Ch.\ Raab, F.\ Schmidt-Kaler, and R.\ Blatt, Nature {\bf 413}, 495
(2001).

\bibitem{Juergen03}
P.\ Buschev, A.\ Wilson, J.\ Eschner, Ch.\ Raab, F.\ Schmidt-Kaler, Ch.\
Becher, R.\ Blatt, Phys.\ Rev.\ Lett.\ {\bf 92}, 223602
(2004).


\bibitem{Alsing92}
P.M.\ Alsing, D.A.\ Cardimona, H.J.\ Carmichael, Phys.\ Rev.\ A {\bf 45}, 1793
(1992).

\bibitem{Alsing92a}
P.\ Alsing, D.-S.\ Guo, H.J.\ Carmichael, Phys.\ Rev.\ A {\bf 45}, 5135 (1992).

\bibitem{Footnote:g=0}
The limit $g(x)=0$ corresponds to no coupling to the cavity where the dynamics
discussed in~\protect\cite{Alsing92} do not apply. One may however ask what
happens when $g(x)\to 0$, where it may seem that the average energy inside the
cavity diverges. In this case, in a lossless resonator
energy would be continuously pumped into the system. These considerations are
not relevant in realistic cases, where cavity decay cannot be neglected.
In any case, when the atom is at a node of the cavity mode it scatters laser
photons and fluorescence is observed.

\bibitem{Footnote:1}
More precisely, the perturbative expansion is valid when
$\kappa,\tilde{\kappa}\ll |{\rm Im}\{\lambda_{\pm}\}|$, where $\lambda_{\pm}$
are the eigenvalues of $H_{\rm eff}$ in Eq.~(\ref{H:eff}) and
$\tilde{\kappa}=\kappa|\beta|^2$. Here,
$|\beta|^2=\Omega^2/g^2$ gives the average number of photons at steady state
in the ideal case $\kappa=0$. The scaling of the perturbative corrections with
$\tilde{\kappa}$ is visible in Eqs.\ (\protect\ref{Res:1}),
(\protect\ref{Res:2}).

\bibitem{Carmichael}
H.J.\ Carmichael, {\it An Open Systems Approach to Quantum Optics},
Springer-Verlag (Berlin, Heidelberg, New York, 1993).

\bibitem{Kimble94}
H.J.\ Kimble, in {\it Cavity Quantum Electrodynamics}, p. 203, ed.\ by P.R.\
Berman, Academic Press (New York, 1994).

\bibitem{Brecha99}
R.J.\ Brecha, P.R.\ Rice, M.\ Xiao, Phys.\ Rev.\ A {\bf 59}, 2392 (1999).

\bibitem{Lounis92}
B.\ Lounis, C.\ Cohen-Tannoudij, J.\ de Phys.\ II (France) {\bf 2}, 579
(1992).

\bibitem{Rice96}
P.R.\ Rice, R.J.\ Brecha, Opt.\ Comm.\ {\bf 126}, 230 (1996).

\bibitem{Domokos02}
P.\ Domokos and H.\ Ritsch, Phys.\ Rev.\ Lett.\ {\bf 89}, 253003 (2002).

\bibitem{Domokos01}
P.\ Domokos, P.\ Horak, H.\ Ritsch, J.\ Phys.\ B {\bf 34}, 187 (2001).


\bibitem{RitschReview}
P.\ Domokos and H.\ Ritch, J.\ Opt.\ Soc.\ Am.\ B {\bf 20}, 1098 (2003).

\bibitem{Zippilli04}
S.\ Zippilli, J.\ Asboth, G.\ Morigi, H.\ Ritsch, to be published in Appl. Phys. B.

\bibitem{OpticalBistability}
R.\ Bonifacio and L.A.\ Lugiato, Opt.\ Comm.\ {\bf 19}, 172 (1976);
Phys.\ Rev.\ A {\bf 18}, 1129 (1978);
L.A.\ Lugiato in {\it Progress in Optics} vol.\ XXI, pp.\ 69ff,
ed.\ E.\ Wolf (North-Holland, Amsterdam 1984).


\bibitem{CARL}
R.\ Bonifacio and L.\ DeSalvo, Nucl.\ Instrum.\ Methods {\bf 341}, 360 (1994);
R.\ Bonifacio, L.\ De Salvo, L.M.\ Narducci, and E.J.\ D'Angelo,
Phys.\ Rev.\ A {\bf 50}, 1716 (1994).

\bibitem{Clemens00}
J.P.\ Clemens and P.R.\ Rice, Phys.\ Rev.\ A {\bf 61}, 063810 (2000).

\bibitem{Nussenzveig01}
C.L.\ Garrido Alzar, M.A.G.\ Martinez, P.\ Nussenzveig, Am.\ J.\ Phys.\ vol.\
{\bf 70} (1), pp. 37 -- 41 (2002).

\bibitem{EIT}
E.~Arimondo, {\it Progress in Optics XXXV}, ed. by E.~Wolf (North-Holland,
Amsterdam, 1996), p. 259; S.E.~Harris, Phys. Today {\bf 50}, No.\ 7, 36
(1997).


\bibitem{Mancini}
S.\ Mancini, D.\ Vitali, P.\ Tombesi,  Phys.\ Rev.\ Lett.\ {\bf 80}, 688
(1998).

\bibitem{FeedbackRempe}
T.\ Fischer, P.\ Maunz, P.W.H.\ Pinkse, T.\ Puppe, G.\ Rempe, Phys.\ Rev.\
Lett.\ {\bf 88}, 163002 (2002).


\end{thebibliography}
\end{document}